\documentclass[aaspp4,11pt,preprint,apjfonts]{emulateapj}

\slugcomment{Accepted for publication in ApJL}
\shorttitle{Suppressed Star Formation in $z\sim2.3$ Galaxies}
\shortauthors{Kriek et al.}

\newcommand{\ha}{H$\alpha$}
\newcommand{\wha}{$W_{\rm H\alpha}$}

\begin{document}

\title{Spectroscopic Identification of Massive Galaxies at $z\sim2.3$
  with Strongly Suppressed Star Formation \altaffilmark{1}}

\author{Mariska Kriek\altaffilmark{2,3},
Pieter G. van Dokkum\altaffilmark{3,4}, 
Marijn Franx\altaffilmark{2},  
Ryan Quadri\altaffilmark{3}, 
Eric Gawiser\altaffilmark{3,4,5},
David Herrera\altaffilmark{3,4},
Garth D. Illingworth\altaffilmark{6}, 
Ivo Labb\'e\altaffilmark{7},
Paulina Lira\altaffilmark{8},
Danilo Marchesini\altaffilmark{3}, 
Hans-Walter Rix\altaffilmark{9},
Gregory Rudnick\altaffilmark{10},
Edward N. Taylor\altaffilmark{2},
Sune Toft\altaffilmark{11},
C. Megan Urry\altaffilmark{4},
and Stijn Wuyts\altaffilmark{2}}

%\email{mariska@strw.leidenuniv.nl}

\altaffiltext{1}{Based on observations obtained at the Gemini
  Observatory, which is operated by the Association of Universities for
  Research in Astronomy, Inc., under a cooperative agreement with the
  NSF on behalf of the Gemini partnership.}

\altaffiltext{2}{Leiden Observatory, PO Box 9513, 2300 RA Leiden,
  Netherlands}

\altaffiltext{3}{Department of Astronomy, Yale University, New Haven,
  CT 06520}

\altaffiltext{4}{Yale Center for Astronomy and Astrophysics, Yale
  University, New Haven, CT 06520}

\altaffiltext{5}{NSF Astronomy and Astrophysics Postdoctoral Fellow}

\altaffiltext{6}{UCO/Lick Observatory, University of California, Santa
  Cruz, CA 95064}

\altaffiltext{7}{Carnegie Fellow, Carnegie Observatories, 813 Santa
  Barbara Street, Pasadena, CA 91101}

\altaffiltext{8}{Departamento de Astronom{\'i}a, Universidad de Chile, 
  Casilla 36-D, Santiago, Chile}

\altaffiltext{9}{Max-Planck-Institute f\"ur Astronomie, K\"onigstuhl
  17, Heidelberg, Germany}

\altaffiltext{10}{Goldberg Fellow, National Optical Astronomy Observatory, 
  950 North Cherry Avenue, Tucson, AZ 85719}
 
\altaffiltext{11}{European Southern Observatory,
  Karl-Schwarzschild-Str. 2, 85748 Garching bei M\"unchen, Germany}

\begin{abstract} 
  We present first results of a spectroscopic survey targeting
  $K$-selected galaxies at $z=2.0-2.7$ using the GNIRS instrument on
  Gemini-South. We obtained near-infrared spectra with a wavelength
  coverage of 1.0--2.5~$\rm\mu$m for 26 $K$-bright galaxies ($K<19.7$)
  selected from the MUSYC survey using photometric redshifts. We
  successfully derived spectroscopic redshifts for all 26 galaxies
  using rest-frame optical emission lines or the redshifted
  Balmer/4000\AA\ break. Twenty galaxies have spectroscopic redshifts
  in the range $2.0<z<2.7$, for which bright emission lines like \ha\
  and [O\,{\sc iii}] fall in atmospheric windows. Surprisingly, we
  detected no emission lines for nine of these 20 galaxies. The median
  $2\sigma$ upper limit on the rest-frame equivalent width of \ha\ for
  these nine galaxies is $\sim$10\,\AA. The stellar continuum emission
  of these same nine galaxies is best fitted by evolved stellar
  population models. The best-fit star formation rate (SFR) is zero
  for five out of nine galaxies, and consistent with zero within
  $1\sigma$ for the remaining four.  Thus, both the \ha\ measurements
  and the independent stellar continuum modeling imply that 45\% of
  our $K$-selected galaxies are not forming stars intensely. This high
  fraction of galaxies without detected line emission and low SFRs may
  imply that the suppression of star formation in massive galaxies
  occurs at higher redshift than is predicted by current CDM galaxy
  formation models. However, obscured star formation may have been
  missed, and deep mid-infrared imaging is needed to clarify this
  situation.
\end{abstract}


\begin{figure*}
  \epsscale{1.15}
  %\epsscale{1.}
  \plotone{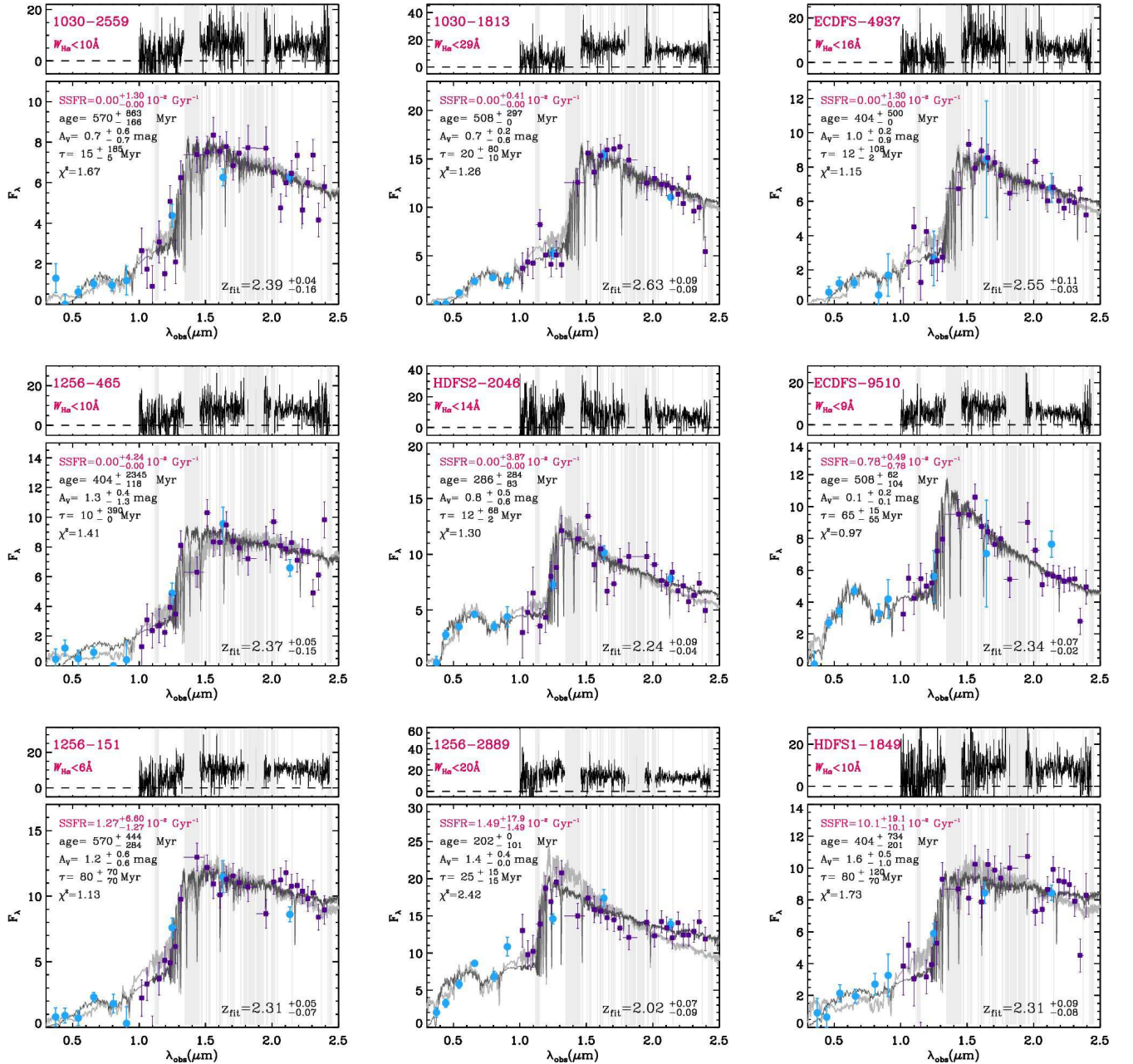}
  \caption{NIR spectra ({\it purple}) and optical-to-NIR photometry
    ({\it blue}) of the nine galaxies at $2.0<z<2.7$ for which we
    detected no \ha\ emission. The upper panels show the 1D original
    spectrum. The `low resolution' binned spectra (160 \AA\ per bin)
    and the photometry are presented in the lower panels. All fluxes
    are given in $10^{-19} \rm ergs~s^{-1}~cm^{-2}~\AA^{-1}$. Regions
    with low or variable atmospheric transmission or with strong sky
    line emission are indicated in light gray. The best fit to the
    optical photometry and low resolution spectrum, allowing all
    values for $A_V$, is overplotted in dark gray. The best-fit model
    parameters are printed in each panel. The corresponding errors are
    the 68\% confidence intervals, derived from 200 Monte Carlo
    simulations. The $\chi^2$ values are given per degree of freedom
    ($N_{\rm deg}=28$). Models without dust generally provide good
    fits as well ({\it light gray}), and imply ages that are about a
    factor of 2 higher than the listed ages. All galaxies without
    detected \ha\ emission are best fitted by evolved stellar
    population models with low specific SFRs.
    \label{spectra}}
\end{figure*}

\begin{figure}
  \begin{center}
    \epsscale{1.15}
    %\epsscale{0.6}
    \plotone{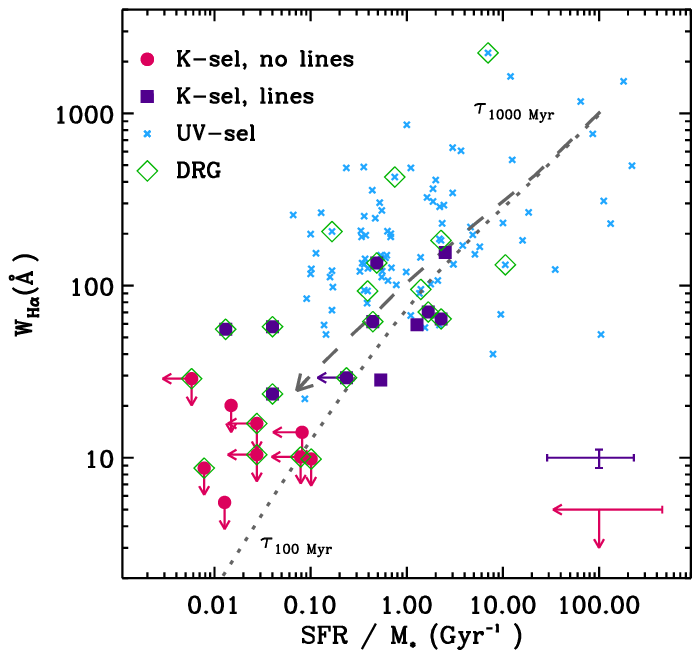}
    \caption{The equivalent width of \ha\ vs. the specific SFR derived
      from our model fits to the spectra. Filled red circles are
      galaxies with no detected \ha\ emission and purple squares are
      emission line galaxies in our sample. Blue crosses are
      UV-selected galaxies by \cite{er06a,er06b,er06c}. DRGs are
      indicated by open green diamonds. Upper limits for \ha\ and the
      specific SFR are 2$\sigma$. Expected relations between \wha\ and
      specific SFR are derived from the \cite{bc03} and \cite{ke98}
      models and drawn for a $\tau_{\rm 100 Myr}$ ({\it gray dotted
        line}) and a $\tau_{\rm 1000 Myr}$ ({\it gray dashed line})
      model for the first 3 Gyr. Both \ha\ measurements and the
      independent SED modeling demonstrate that in 9 out of 20
      galaxies in our sample the star formation has been strongly
      suppressed.\label{wha}}
  \end{center}
\end{figure}

\section{INTRODUCTION}

Observations imply that stellar populations of high-mass galaxies were
formed at higher redshift than those of low-mass galaxies
\citep[e.g.,][]{co96,ju05}. Recent hierarchical CDM models are able to
produce dead, massive galaxies at high redshift by incorporating
feed-back from active galactic nuclei \citep[AGN;
e.g.,][]{cr06,bo06,dl06,ho06}. In order to observationally determine
when and how star formation in massive galaxies was suppressed it is
necessary to identify and study these objects out to the highest
redshifts.

Recently, massive and apparently dead galaxies have been identified at
$z>1.5$ \citep[e.g.,][]{mc04,sa05,da05,la05,re05,re06,pa06}. Beyond
$z=2$ most studies rely on photometric redshifts and broad-band colors
to identify these galaxies. However, as dust and age have similar
effects on the broad-band spectral energy distribution (SED), the star
formation rates (SFRs) are often not well constrained. Furthermore,
their photometric redshifts are not well calibrated, as only a few at
$z>2$ have spectroscopic redshifts \citep{da05}. Thus spectroscopic
redshifts and independent stellar population diagnostics are needed to
determine the prevalence of ``red and dead'' galaxies beyond $z=2$.

Spectroscopic confirmation of non-star-forming galaxies at $z>2$ is
complicated due to their faint rest-frame UV emission and lack of
nebular emission lines. Deep near-infrared (NIR) spectroscopy provides
the best option to confirm such galaxies at $z>2$, as their relatively
bright rest-frame optical luminosity allows direct detection of the
stellar continuum. The optical continuum shape, and in particular the
Balmer/4000 \AA\ break, can be used to derive redshifts for galaxies
without emission lines, and provides independent constraints on
stellar populations \citep{kr06}.  

To study a high-redshift spectroscopic sample that is not biased
towards galaxies with bright emission lines, we are conducting a NIR
spectroscopic survey of $K$-selected galaxies with photometric
redshifts $z\sim2.3$. Here we report on a surprising result of our
survey: the large fraction of galaxies with no detected emission
lines. Throughout the letter we assume a $\Lambda$CDM cosmology with
$\Omega_{\rm m}=0.3$, $\Omega_{\rm \Lambda}=0.7$, and $H_{\rm
  0}=70$~km s$^{-1}$ Mpc$^{-1}$. All broadband magnitudes are given in
the Vega-based photometric system.

\section{GALAXY SAMPLE AND DATA}

The galaxies presented in this letter are drawn from the
Multi-wavelength Survey by Yale-Chile (MUSYC). This survey provides
optical and deep NIR photometry for several southern and equatorial
fields \citep{ga06,qu06}. The targets were selected in $K$ ($K<19.7$)
to reduce the dispersion in stellar mass and to ensure adequate $S/N$
in the NIR spectra. Additionally, we required a photometric redshift
in the range $2.0<z<2.7$, for which bright rest-frame optical emission
lines such as [O\,{\sc iii}] and \ha\ fall in the $H$ and $K$
atmospheric windows. The photometric redshifts are derived following
the procedure described in \cite{ru01,ru03}.

We observed 26 galaxies with the Gemini near-infrared spectrograph
(GNIRS) in 2004 September (GS-2004B-Q-38), 2005 May (GS-2005A-Q-20),
2006 January (GS-2005B-C-12) and 2006 February (GS-2006A-C-6). The
spectra of two of these galaxies have already been presented by
\cite{vd05} and \cite{kr06}. All galaxies were observed in
cross-dispersed mode, in combination with the short wavelength camera,
the 32 line mm$^{-1}$ grating (R=1000) and the 0\farcs675 by 6\farcs2
slit. In this configuration we obtained a wavelength coverage of 1.0
-- 2.5 $\rm \mu m$. The galaxies were observed for 1-4 hours,
depending on the brightness of the target and the weather conditions.
The observational techniques and reduction of the GNIRS spectra are
described in detail by \cite{kr06}. For each galaxy we extract a
one-dimensional original and low resolution binned spectrum.

To derive the stellar population properties and obtain redshifts for
galaxies without emission lines, we fit stellar population models to
the low resolution continuum spectra together with the $UBVRIz$
fluxes, following the technique described in \cite{kr06}.  We use the
\cite{bc03} models with a set of exponentially declining star
formation histories, a \cite{sa55} initial mass function (IMF) between
0.1 and 100 $M_{\odot}$, and solar metallicity, and adopt the
\cite{ca00} reddening law. The assumed model parameters (IMF,
reddening law, metallicity) are identical to those used by e.g.,
\cite{fo04}, \cite{sh05}, and \cite{pa06}.

We obtained spectroscopic redshifts for all 26 galaxies using
rest-frame optical emission lines or the Balmer/4000\AA\ break. The
`break' redshifts have a median uncertainty of $|\Delta
z|/(1+z)=0.017$, as determined from fitting the low-resolution
continua of emission line galaxies with $z$ as a free parameter. 20 of
26 galaxies have spectroscopic redshifts in the range $2.0<z<2.7$, for
which \ha\ falls in the $K$-band. In what follows we restrict the
sample to the galaxies in this redshift range; the full sample will be
described elsewhere (M.~Kriek et al., in preparation). We note that
the six galaxies that fall out of this redshift range have
$z=1.75-1.95$.\footnote{The median of the differences between
  spectroscopic and photometric redshifts $(z_{\rm spec}-z_{\rm
    phot})/(1+z_{\rm spec})$ is only $-0.001$ for the full sample of
  26 galaxies.}

\section{SUPPRESSED STAR FORMATION}

Surprisingly, nine out of the 20 galaxies in the sample show no
emission lines in their rest-frame optical spectra. The spectra and
best fits of these galaxies are presented in Fig.~\ref{spectra}. For
these galaxies we derived upper limits on the \ha equivalent width
(\wha) as follows. We have drawn 200 random redshifts from the
redshift probability distribution, and determined the 2$\sigma$ upper
limit of \wha\ in each case from the measured noise properties,
assuming a rest-frame \ha\ FWHM of 500 km s$^{-1}$ \citep[see][]{vd04}
and the best-fit stellar continuum. The adopted limit is the maximum
value found in the simulations, excluding the highest 5\%. The median
\wha\ upper limit of these nine galaxies (corrected for Balmer
absorption) is 10\,\AA.

\wha\ is a measure of the ratio of current to past star formation, and
the limits on \wha\ may imply very low SFRs in these galaxies. We
investigate this in Fig.~\ref{wha}, in which we plot the rest-frame
\wha\ upper limits (corrected for Balmer absorption) versus the
specific SFRs (SFR / $M_*$) derived from our model fits to the
spectra. Remarkably, all nine galaxies without detected emission
lines ({\it filled red circles}) are best fitted by stellar population
models with low specific SFRs, and the data points are broadly
consistent with the expected relations between these properties
\citep{ke98,bc03}. For five galaxies the best-fit SFR is zero and the
four remaining galaxies have best-fit values that are consistent with
zero within $1\sigma$. The midmean\footnote{mean of the central two
  quadrants} specific and absolute SFRs are $0.004\rm\,Gyr^{-1}$ and
$0.9\,M_{\odot}/$yr respectively, significantly lower than the
$0.56\rm\,Gyr^{-1}$ and $128\,M_{\odot}/$yr found for the eleven
emission line galaxies.  Thus, both the \ha\ measurements and the
stellar continuum modeling imply that the star formation in these nine
galaxies has been strongly suppressed. We note that we find a similar
relation when we plot $L_{\rm H\alpha}$ vs.\ the modeled absolute SFR,
as our galaxies span only a small range in stellar mass
($0.9-4.6\,\times\,10^{11}M_{\odot}$). These nine galaxies have a
median stellar mass of $2.6\times 10^{11} M_{\odot}$, a median $J-K$
color of 2.45, and have undergone a median of 21 age/$\tau$ e-folding
times. Six out of nine are distant red galaxies \citep[DRGs,
$J-K>2.3$, ][]{fr03}. As can be seen in Fig.~\ref{wha} several of the
emission line galaxies are also best fitted by stellar population
models with low specific SFRs. This may suggest that the gas in these
galaxies is not ionized by hot stars; we will explore this in a future
paper.

Formally, we find high best-fit values for the dust content for eight
out of nine galaxies without emission lines. However, $A_V$ is poorly
constrained for most of these galaxies, and models with zero or only
small amounts of dust are consistent within $1\sigma$. We refitted all
9 galaxies allowing only models without dust. The best fits
(Fig.~\ref{spectra}, {\it light gray}) also yield low specific SFRs,
ranging from $0.001-0.013\rm\,Gyr^{-1}$, with a median value of
$0.002\rm\,Gyr^{-1}$. The median best-fit stellar age is 0.9 Gyr,
which is a factor of $\sim2$ larger than when allowing dust. While we
cannot draw firm conclusions about the dust content in these galaxies,
we note that high $A_V$ may indicate that they are still in the
process of losing their gas and dust.

Figure~\ref{wha} also shows the UV-selected $z\sim2$ galaxies of
\cite{er06a,er06b,er06c}. These galaxies have similar \wha\ and
specific SFRs as the $K$-selected galaxies with emission
lines. However, there is no overlap with the $K$-selected galaxies
without detected \ha\ emission; both the \wha\ and the modeled
specific SFR are higher for the UV-selected galaxies. Although
UV-selection is able to find massive galaxies
\citep[e.g.,][]{sh05,er06b}, it does not sample the full distribution
of their properties \citep[see also][]{vd06}.

\section{DISCUSSION}

We find that nine out of 20 $K$-selected galaxies at $2.0<z_{\rm
  spec}<2.7$ have no detected \ha\ emission (\wha\,$\lesssim$~10\AA)
and are best fit by stellar population models with low specific SFRs
($\sim\,0.004\rm\,Gyr^{-1}$), implying a fraction of galaxies with
strongly suppressed star formation of $45^{+18}_{-12}$\,\%.  The
quoted uncertainty is derived assuming Poisson statistics, and does
not include the following systematic errors and caveats. First, our
$z_{\rm phot}$ selection criterion could introduce biases, as
systematic errors in photometric redshift may correlate with SED
type\footnote{This effect may be small, as the distribution of
  specific SFRs of the six objects at $z_{\rm spec}<2$ is similar to
  that of the 20 remaining objects}. Also, the $K$-band selection
criterion could bias our sample, as for starburst galaxies strong
emission lines can contribute significantly to the $K$-band flux
\citep{er06a}. We do not expect this bias to be strong, as the median
contribution of the emission lines to the $K$-band is only 0.04 mag
for the eleven emission line galaxies. Third, incompleteness may play
a role, as we observed only $\sim$20\% of the galaxies that meet the
selection criteria. We note, however, that according to RS- and
KS-tests our $K$-selected sample has a similar distribution of
rest-frame $U-V$-colors as the large mass-limited sample ($>10^{11}
M_{\odot}$) by \cite{vd06}, when applying the same $K$-magnitude
cut. Furthermore, the assumption that \ha\ emission is only due to
star formation may lead us to underestimate the fraction.

Finally, and perhaps most importantly, we may miss star formation with
very high rest-frame optical extinction. Although most local ULIRGs
have high integrated \wha\ \citep[$W_{\rm H\alpha+[N\,{\sc ii}]} \sim
90$\AA,][]{li95}, for some the star-burst regions are almost
completely obscured (e.g., Arp 220 has $W_{\rm H\alpha+[N\,{\sc ii}]}
= 18$\AA) and these objects might be misinterpreted in our
analysis. Available 24 $\mu$m imaging on DRG samples shows that
30-50\% have no MIPS counterpart \citep{we06,pa06,re06}, and
\citet{re06} finds that red galaxies with low MIPS fluxes typically
have low specific SFRs. To resolve this issue it is necessary to
combine our spectra with deep mid/far-infrared imaging to detect
hidden star formation.
  
It is interesting to compare our fraction of galaxies with strongly
suppressed star formation to previous studies\footnote{The fractions
  of galaxies with suppressed star formation among DRGs and
  $K$-selected galaxies are similar in our sample}. Our result is
consistent with the fraction found by \cite{la05}, as they identified
three ``red and dead'' galaxies from a sample of 11 DRGs in the HDF-S,
and with the study of \cite{re06}, who find that seven out of 24 DRGs
at $1.5<z<2.6$ are not detected at 24 $\mu$m, and have an average low
specific SFR of 0.05 Gyr$^{-1}$. However, our fraction is
significantly higher than the fractions found by \cite{da04} and
\cite{pa06}. Using the $BzK$ criterion, \cite{da04} identify all their
11 $K$-selected galaxies at $2.0<z<2.7$ as star-forming galaxies. The
difference may be partly explained by different definitions, as
$\sim$4 of our 9 galaxies with suppressed star formation galaxies
would have been identified as star-forming galaxies by the $BzK$
criterion. \cite{pa06} finds that $\sim$10\% of 153 DRGs at $1.5\le
z_{\rm phot}\le 3$ show no signs of current star formation. Again, the
selection criteria could play a role; none of our galaxies would have
been classified as ``dead'' by \cite{pa06}, as they apply the
following criteria: age $>1 \rm Gyr$ , $E(B-V)\le0.1$, age$/\tau>3$
and no X-ray or 24$\mu$m detection. The sample selection could also be
a factor, as the contribution of dusty star-forming galaxies to the
DRG population is expected to be higher at $z<2$. Furthermore, the
CDF-S field -- in which both the studies by \cite{pa06} and
\cite{da04} were performed -- may be atypical \citep{vd06}. We stress
that our study is the first that is based on spectroscopic redshifts,
and that this might also account for differences in the obtained
fractions. This will be explored in a future paper.

We note that most current CDM galaxy formation models fail to produce
the high fraction of red galaxies at the massive end as found by
\cite{vd06}. One easy way to solve this is by allowing more dust in
the galaxies, but we have shown here that a large fraction of galaxies
has low specific SFRs, and these are generally absent in these models
at $z>2$ \citep[e.g.,][]{so04,na05,ka06}. Our results may indicate
that the suppression of the star formation in the most massive
galaxies occurs at higher redshift than has been predicted by current
models. In this context, it is interesting to note that not all of the
line-emission in our sample may be due to star formation: from our
low-resolution GNIRS spectra it appears that several of the
emission-line galaxies with low specific SFRs exhibit high [N\,{\sc
  ii}]/H$\alpha$ ratios, possibly indicating AGN. High resolution
spectra of the line-emitting objects in our sample will be discussed
in a future paper (M.~Kriek et al., in preparation).

\acknowledgments We thank the referee for constructive comments which
improved the manuscript. This research was supported by grants from
the Netherlands Foundation for Research (NWO), the Leids
Kerkhoven-Bosscha Fonds, National Science Foundation grant NSF CAREER
AST-044967, and NASA LTSA NNG04GE12G.

\end{document}